\documentclass[twocolumn,showpacs,preprintnumbers]{revtex4}
\usepackage{graphicx}
\usepackage{bm}
\usepackage{color}
\usepackage[normalem]{ulem}

\graphicspath{%
    {converted_graphics/}
    {/}
}
\begin{document}

\title{Decoherence by spontaneous emission: a single-atom analog of superradiance}

\author{Reinaldo de Melo e Souza, Fran\c{c}ois Impens, Paulo A. Maia Neto}
\date{\today}

\affiliation{Instituto de F\'{i}sica, Universidade Federal do Rio de Janeiro,  Rio de Janeiro, RJ 21941-972, Brazil }

\begin{abstract}
We 
show that the decoherence of the atomic center-of-mass induced by spontaneous emission involves interferences
corresponding to 
a single-atom analog
of superradiance.
We use a decomposition of the stationary decoherence rate as a a sum of local and nonlocal contributions obtained  to second order in the interaction by the influence functional method.  These terms are respectively related to the strength of the coupling between system and environment, and to the quality of the information about the system leaking into the environment.  While the local contribution always yields a positive decoherence rate, the nonlocal one
  may lead to recoherence when only partial information about the system is obtained from the disturbed environment. 
  The nonlocal contribution contains interferences between different quantum amplitudes leading to oscillations of the decoherence rate reminiscent of superradiance. 
   These concepts, illustrated here in the framework of atom interferometry within a trap, may be applied to a variety of quantum systems.
\end{abstract}

\maketitle

\section{Introduction}

Decoherence is a fundamental issue of quantum mechanics that has been the object of innumerous theoretical and experimental studies~\cite{ZurekRMP03}.  
The dynamics of open quantum systems is particularly intriguing in the case of 
  non-Markovian environments~\cite{Breuer09,BreuerRMP16,Lombardo15,Groeblacher15,Genkin16,Semin16}. {  In this paper, we show that an environment of finite memory enables a quantum system with a single atom to mimic collective effects such as superradiance~\cite{Dicke54}, which is still a topic of intense investigation~\cite{Bachelard12,Ritsch12,Oliveira14,Wang15,Kaiser16b,Stobbe16,Zanthier16,Damanet16}.

The influence functional~\cite{Feynman1963,CalzettaHuBook} is an effective theoretical tool to capture the effect  of a non-Markovian environment on a quantum system. In this framework, the finite memory time of the environment induces a coupling between the backward and forward histories of the system within the closed-time-path~(CTP) integrals describing the density matrix evolution.
Such coupling leads to a  nonlocal functional, which cannot be split into separate contributions from forward and backward histories. When the environment has a finite memory, this provides a contribution of crucial importance to decoherence, which may be negative (i.e. recoherence in the sense defined by Ref.~\cite{Mazzitelli2003}). 
  The nonlocal contribution to the decoherence rate exhibits interference oscillations which constitute an analog of superradiance~\cite{Dicke54}.  
  
The paper is organized as follows. In Sec.~II, we discuss the coupling between backward and forward histories in the  closed-time path formalism and its relation with nonlocality.  An explicit perturbative expansion is developed in Sec.~III. 
 Sec.~IV introduces an alternative approach enabling the connection with the quality of which-path information leaking into the environment. 
In Sec.~V, we show that the decoherence of a single atom oscillates as a function of the distance between the wavepackets, which is reminiscent of superradiance.
 Concluding remarks are presented in Sec.~VI.

\section{Coupling between backward and forward histories}

 Precisely, we consider a system described by a center-of-mass position $\mathbf{r}$ interacting with a generic environment.  It is well-known that the evolution of pure quantum states can be described in terms of forward path integrals. 
 On the other hand, an open quantum system can be suitably described 
  by a density matrix, whose  elements
  contain products of wave functions by their complex conjugates.
 Therefore, it is possible to write the evolution of a density matrix as a double path-integral, with the complex conjugation representing a backward propagation~ \cite{Feynman1963,CalzettaHuBook,MiltonBook15, MahanBook}.  The reduced  density matrix of our system is propagated with the CTP integral   
\begin{eqnarray}
\label{eq:CTP integral atomic position}
{ \langle \mathbf{r}_f | \rho(\Delta t)  | \mathbf{r}_f' \rangle} =  \int_{\rm CTP}^{\mathbf{r}_f,\mathbf{r}_f'} \!\! \mathcal D \mathbf{r}  \:  e^{\frac {i} {\hbar} \left( S_0[\mathbf{r}] - S_0[\mathbf{r}'] + S_{\rm IF}[\mathbf{r},\mathbf{r}'] \right)}
\end{eqnarray}
where we used the following notation 
\begin{equation}
\label{eq:CTP notation}
 \int_{\rm CTP}^{\mathbf{r}_f,\mathbf{r}'_f}  \! \! \!  \mathcal D \mathbf{r} \!  = \!  \int \! d \mathbf{r}_0 d \mathbf{r}'_0 \! \! \int_{\mathbf{r}(0)=\mathbf{r}_0}^{\mathbf{r}(\Delta t)=\mathbf{r}_f} \! \! \! \! \mathcal D \mathbf{r} \! \int_{\mathbf{r}'(0)=\mathbf{r}'_0}^{\mathbf{r}'(\Delta t)=r'_f}  \! \! \! \!  \mathcal D \mathbf{r}'   { \langle \mathbf{r}_0 | 
 \rho (0) | \mathbf{r}_0' \rangle}
\end{equation}
 with $\rho(0)$ the initial system density matrix  and $S_0[\mathbf{r}]$ the interaction-free action of the system, i.e. when all couplings to the other
 degrees of freedom
  (d.o.f.s) - referred to as the environment - are ignored. 
 We have assumed that  system and environment are initially uncorrelated. 
 As the actions $S_0[\mathbf{r}]$ and $S_0[\mathbf{r}']$  appear with different signs, the paths $[\mathbf{r}(t)]$ and $[\mathbf{r}'(t)]$ correspond respectively to forward and backward histories. The influence of the environment on the system is completely described by the complex influence functional $S_{\rm IF}[\mathbf{r},\mathbf{r}'] $, which can be decomposed into a sum of local and nonlocal terms, namely $S_{\rm IF}[\mathbf{r},\mathbf{r}']=S_{\rm L}[\mathbf{r},\mathbf{r}']+S_{\rm NL}[\mathbf{r},\mathbf{r}'].$ The local influence functional is expressed 
  in terms of  the single 
 functional $S_{\rm single}$ depending on one path at a time:  $S_{\rm L}[\mathbf{r},\mathbf{r}']=S_{\rm single}[\mathbf{r}]-S_{\rm single}^*[\mathbf{r}']$
 (explicit expressions are given below). 
  In contrast, the nonlocal influence functional $S_{\rm NL}[\mathbf{r},\mathbf{r}']$ cannot be reduced to such a combination of single-path functionals.  
 The presence of this nonlocal functional couples the backward and forward histories, as illustrated in Figure~\ref{fig:CTPdiagrams}. This coupling occurs over a time scale of the order of the memory time of the environment.
 
\begin{figure}[htbp]
\begin{center}
\includegraphics[width=8.5cm]{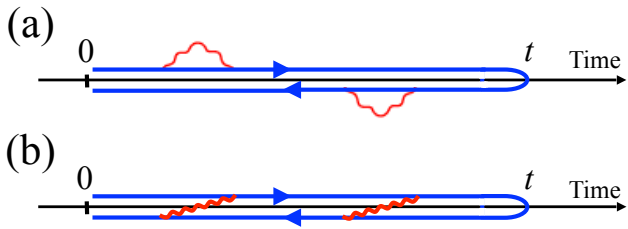}
\end{center}  \caption{(color online). 
  Closed-time-path diagrams representing (a) local  and (b) nonlocal contributions to the 
 influence functional. The nonlocal influence terms couple the backward and forward histories.}
\label{fig:CTPdiagrams}
\end{figure} 

The real part of the influence functional describes a relative phase shift  between the two paths, while its imaginary part, henceforth denoted by $S_{\rm L(NL)}^{\rm Dec}[ \mathbf{r},\mathbf{r}'] = \mbox{Im} \left[   S_{\rm L(NL)}[ \mathbf{r},\mathbf{r}'] \right]$, provides the exponential attenuation of the interference fringes~\cite{BookBreuerPetrucione}. Note that the expression of the local influence functional $S_{\rm L}[\mathbf{r},\mathbf{r}']$ as a difference between a functional and its complex conjugate is perfectly consistent with this interpretation. Indeed, this form ensures that its real and imaginary parts are respectively antisymmetric and symmetric under the exchange of paths. These features are actually needed to preserve the hermiticity of the reduced density matrix $\rho(\mathbf{r}_f,\mathbf{r}_f';t )=\rho^*(\mathbf{r}_f',\mathbf{r}_f;t )$ during the propagation. As a consequence, the local decoherence rates associated to 
 $S_{\rm L}^{\rm Dec}[ \mathbf{r},\mathbf{r}']$
 sum up.

 If one considers only the decoherence resulting from the local influence functional, one obtains results in contradiction with the common interpretation of decoherence as a flow of which-way information from the system to its environment. Let us assume that the system is put in a coherent superposition state of paths $[\mathbf{r}_1(t)]$ and $[\mathbf{r}_2(t)]$. For simplicity, we assume that the density matrix 
 corresponds to thin wavepackets, so that only the quantum paths lying in a tight bundle around the central  paths $[\mathbf{r}_1(t)],[\mathbf{r}_2(t)]$ significantly contribute to the integrals~(\ref{eq:CTP integral atomic position},\ref{eq:CTP notation}). Should these two paths become arbitrarily close to one another, and making the reasonable assumption that the
 single-path functional $S_{\rm single}[\mathbf{r}]$ is continuous, one would get from 
 $S_{\rm L}^{\rm Dec}[\mathbf{r}_1(t),\mathbf{r}_2(t)]=\mbox{Im}(S_{\rm single}[\mathbf{r}_1(t)]+S_{\rm single}[\mathbf{r}_2(t)])$ a system decoherence rate which is twice the contribution arising from each path. On the other hand, in this limit, the two paths are so close that no matter each way the system goes, it leaves an almost identical record on the environment. Thus,  the amount of which-way information flowing to the environment gradually vanishes, but the local decoherence rate remains finite (similar conclusions hold for wavepackets of finite width).
  
 This example shows that the nonlocal influence functional must be taken into account. Its contribution to decoherence, given by
$S_{\rm NL}^{\rm Dec}[\mathbf{r},\mathbf{r}'],$ captures the distinguishability between the 
two  paths $[\mathbf{r}_1(t)]$ and $[\mathbf{r}_2(t)].$ 
This property requires a functional involving simultaneously both paths at a time, i.e. a nonlocal functional. In the limit above, when the two paths $[\mathbf{r}_1(t)]$ and $[\mathbf{r}_2(t)]$ become identical, this contribution should cancel exactly the local contribution from $S_{\rm L}[\mathbf{r},\mathbf{r}']$, enabling one to retrieve the expected absence of decoherence. Thus, in this limit, the nonlocal influence functional should provide complete recoherence to the system. The derivation below confirms this argument: indeed the nonlocal contribution to decoherence has then the same amplitude as the local decoherence, and opposite sign. 

The opposite limit is reached when the paths $[\mathbf{r}_1(t)]$ and $[\mathbf{r}_2(t)]$ are so different that, when followed by the system, their respective perturbations of the environment correspond to orthogonal quantum states. In other words, perfect which-way information flows from the system to the environment: by performing a suitable measurement of the perturbation suffered by the environment, an observer could tell with certainty which path was taken by the system. As established below, the nonlocal decoherence contribution vanishes in this regime, meaning that the decoherence saturates as a function of the distance between the paths. 
More generally,  the total decoherence rate results from a fine-tuning between a local decoherence rate of positive sign associated to the probability of a physical process, and a nonlocal decoherence rate of indefinite sign, to be interpreted as a recoherence~\cite{Mazzitelli2003} rate when bearing a negative sign. This nonlocal recoherence measures the lack of which-way information obtained from the system-environment interaction process.\\

It has been shown previously that the environment may induce, through the influence functional, nonlocal phase shifts in systems such as atom interferometers when considering the dipolar interaction~\cite{Impens13a,Impens14}. However, in these examples, the nonlocal contribution represents
 only a tiny (relativistic) correction to other local atomic phases induced by the environment. In contrast, as far as decoherence is concerned, the nonlocal influence contribution (depicted in Fig.~\ref{fig:CTPdiagrams}b) 
can be of the same order of magnitude as the local term. 
 By capturing the quality of the which-way information  leaking into the environment, the nonlocal contribution is hence absolutely essential to estimate the decoherence.

\section{Local and nonlocal decoherence functionals}

 We address the decoherence 
effect over a time scale $\Delta t$ simultaneously much larger than the 
 self-dynamics of the
 environment  and much shorter
 than the  time scale associated to the strength of the coupling with the system.
For instance, the natural width of the atomic transition  is several orders of magnitude smaller than the transition frequency in usual atomic states,
allowing us to consider the intermediate time $\Delta t.$
 The average decoherence rate on this time scale will be referred to as the stationary decoherence rate.

 We first use CTP path integrals in order to obtain the local and nonlocal stationary decoherence rates, noted $\Gamma_{\rm L}$ and $\Gamma_{\rm NL}$ respectively, and defined as
\begin{equation}
\label{eq:definition stationary decoherence}
 \Gamma_{\rm L (NL)} = \lim_{\Delta t \rightarrow +\infty} \frac 1 {\hbar \: \Delta t} \left.  S_{\rm L (NL)}^{\rm Dec}[ \mathbf{r}_1(t),\mathbf{r}_2(t)]  \right|_{t \in [0,\Delta t]}
 \end{equation}
 for an initial quantum superposition of thin wavepackets centered at prescribed paths  $[\mathbf{r}_1(t)]$ and $[\mathbf{r}_2(t)]$, defined on the time interval $[0,\Delta t].$

 In order to compute the r.-h.-s. of Eq.~(\ref{eq:definition stationary decoherence}), we assume that the 
 total action has the general form 
 $S[\mathbf{r},\mathbf{q},\mathbf{X}]=S_{0}[\mathbf{r}]+S_{q}[\mathbf{q}]+S_X[\mathbf{X}]+S_{\rm int}[\mathbf{r},\mathbf{q},\mathbf{X}],$
 with $\mathbf{q}$ and $\mathbf{X}$ representing environment d.o.f.s. and $S_{q},$ $S_X$  the corresponding free actions. 
We start by coarse-graining over the fastest environment d.o.f.s., making use of a hierarchy of time-scales 
(see for instance ~\cite{Ryan11}).  Here, we assume that the continuum of d.o.f.s $\mathbf{X}(x)$ fluctuates on a much faster time scale than the d.o.f.s $\mathbf{q}(t).$ 
In the example to be discussed later, $\mathbf{X}(x)$ represents the electromagnetic (EM) field and  $\mathbf{q}(t)$
 the atomic dipole, so that the fluctuations of the former are indeed short-lived on an atomic dipole time-scale~\cite{CohenAtomPhotons}. 
 Thus, we first coarse-grain over the environment d.o.f.s $\mathbf{X}(x)$, yielding an auxiliary influence functional  $S_{\rm IF}[\mathbf{r},\mathbf{r}',\mathbf{q},\mathbf{q}']$, and then trace out the slower environment d.o.f.s $\mathbf{q}(t)$. 
 We consider general free actions and only require   the interaction to be local,
 that is,
\begin{eqnarray}
\label{eq:definition Sint}
S_{\rm int}[\mathbf{X},\mathbf{q},\mathbf{r}]=\int_{0}^{\Delta t} \! \! dt \: V(\mathbf{X}(\mathbf{r},t),\mathbf{q}(t)) \, .
\end{eqnarray}
For simplicity, we also assume very thin wavepackets, since the generalization to more general packets is straightforward \cite{Impens14}.
We trace out the environment degrees of freedom using perturbation
theory and suppose the first order to vanish  (that is $\langle \mathbf{q} (t) \rangle = 0 $ or $\langle \mathbf{X} (\mathbf{r},t) \rangle = 0 $ at any time $t$) \footnote{Such condition is met, for instance, if one takes for the potential ${V}(\mathbf{r})$ the dipolar interaction, and if the atoms are initially in an unpolarized quantum state or in presence of the vacuum field.}.

In  Appendix A, we show that the local and nonlocal influence functionals are then given by 
\begin{eqnarray}
S_{\rm L}[ \mathbf{r}_1(t),\mathbf{r}_2(t)]  \!  \!  &= & \!  \!    \frac {i} {\hbar} \int _{0}^{\Delta t} \!  \! \!  dt \int _{0}^{\Delta t} \! \! \!  dt'   \Theta(t-t')      
  \left(  \langle {\widetilde{V}}(r_1(t))   {\widetilde{V}}(r_1(t')) \rangle \right.\nonumber\\
  \label{eq:phase local standard perturbation}
  &&
  + \left. \langle    {\widetilde{V}}(r_2(t')) { \widetilde{V}}(r_2(t))    \rangle \right)    
\end{eqnarray}
\begin{eqnarray}
&  & S_{\rm NL}[ \mathbf{r}_1(t),\mathbf{r}_2(t)] \!  \! = \!  \!  - \frac {i} {\hbar} \!   \int _{0}^{\Delta t}  \!   \!  \!  dt \int _{0}^{\Delta t}  \!   \! \!   dt' \langle   {\widetilde{V}}(r_2(t))   {\widetilde{V}}(r_1(t'))  \rangle  \nonumber\\  \label{eq:phase nonlocal standard perturbation}
\end{eqnarray}
where 
$\Theta$ represents the Heaviside step function and 
${\widetilde{V}}(r(t))={\widetilde{V}}(\mathbf{r}(t),t) $ denotes
the operator $V$ in the interaction picture obtained from the free
evolution of the operators $\mathbf{X}$ and $\mathbf{q}$. $\langle\cdots\rangle$ denotes the expectation value in the initial state of the environment. 

A situation of particular interest is the one where the d.o.f.s $\mathbf{q}$ and $\mathbf{X}$ are linearly coupled: $V(\mathbf{r},t)=\mathbf{q}(t)\cdot\mathbf{X}(\mathbf{r},t).$ Substituting
this equation into Eqs.~(\ref{eq:phase local standard perturbation},\ref{eq:phase nonlocal standard perturbation}) and taking the imaginary part we obtain

   \begin{eqnarray}
  \label{eq:expressionSL}
  S_{\rm L}^{\rm Dec} [ \mathbf{r}_1(t),\mathbf{r}_2(t)]  &=&\int_0^{\Delta t} \!\! dt \int_0^{\Delta t}  \!\! dt' [G(r_1(t),r_1(t'))\\
  \nonumber
   &&\,+\,G(r_2(t'),r_2(t))]  \\
   \label{eq:expressionSNL}
  S_{\rm NL}^{\rm Dec}[ \mathbf{r}_1(t),\mathbf{r}_2(t)] &=&   \int_0^{\Delta t}  \!\! dt \int_0^{\Delta t} \!\! dt' G(r_1(t),r_2(t'))
         \end{eqnarray} 
where we employed the shorthand notation $r_i(t)=(t,\mathbf{r}_i(t))$ ($i=1,2$) and
\begin{eqnarray}
  \label{eq:expressionG}
G(x,x') & \equiv & \frac {1} {4 \hbar}  \sum_{i,j}  \left[   \Theta(t-t')\langle [ {q}_i(t),  {q}_j(t') ] \rangle  \langle [{X}_i(x), {X}_j(x') ] \rangle  \nonumber \right. \\
&  & +\left.  \langle \{ {q}_i(t),  {q}_j(t') \}  \rangle  \langle \{ {X}_i(x), {X}_j(x') \} \rangle  \right] \, .
\end{eqnarray}
In the sum, the indices $ i,j $ span the Cartesian  components of the vectors $\mathbf{q}$ and $\mathbf{X}(x).$
Combining these expressions with Eq.~(\ref{eq:definition stationary decoherence}), one obtains the stationary decoherence rates 
 $\Gamma_{\rm L}$ and $ \Gamma_{\rm NL}$ predicted by the influence functional method.

The nonlocal contribution to decoherence $S_{\rm NL}^{\rm Dec} [ \mathbf{r}_1(t), \mathbf{r}_2(t)] $ originates from the mixing of the forward $[\mathbf{r}_1(t)]$ and backward $[\mathbf{r}_2(t)]$ histories within the CTP path integral. 
As suggested by
 Eqs.~(\ref{eq:expressionSNL}) and (\ref{eq:expressionG}), 
such coupling can only occur in practice if the memory time of the d.o.f.s $\mathbf{q}(t)$, captured by the decay of the correlation functions  $\langle {q}_i(t) {q}_j(t') \rangle,$ is greater than the  propagation time
 of the interaction connecting the two histories. 
 For instance, cold atoms can provide internal d.o.f.s with sufficiently long-lived correlations
  enabling the connection between forward and backward histories  at distinct locations. In this sense, cold atomic systems provide an adequate playground for probing non-Markovian environments for open quantum systems.

 It is possible to generalize the argument above to  multiple-path atom interferometers~\cite{CroninRMP09}. It turns out that the nonlocal contribution to decoherence involves a sum over all possible pairs of paths, similarly to the non-additive atomic phases induced by electromagnetic vacuum field fluctuations~\cite{Impens13b}.

\section{Nonlocality and quality of information}

To have a complementary insight on the nature of the nonlocal decoherence, we present below an alternative derivation, based
on standard time-dependent perturbation theory.
  As in the previous derivation, we assume that 
  the system and environment are initially uncorrelated, described by a pure quantum state of the form  
 \begin{equation} \label{0}
  | \Psi (0) \rangle = \frac {1} {\sqrt{2}} \left( | \psi_1(0)  \rangle + | \psi_2(0) \rangle \right) \otimes | \psi_{E}(0) \rangle,
  \end{equation}
and that their  coupling  can be treated perturbatively. 
The system quantum states $| \psi_1(0)  \rangle $ and $| \psi_2(0)  \rangle $
  correspond to  propagations  along paths $[\mathbf{r}_1(t)]$ and  $[\mathbf{r}_2(t)],$ respectively, and $| \psi_{E}(0) \rangle$ is the initial state of the environment.
Although we have chosen a particularly simple initial state for the system, the discussion to follow could be generalized without  difficulty to an initial quantum state of the form $| \psi(0) \rangle = \sum_{k=1}^N \alpha_k | \psi_k(0) \rangle  \otimes  | \psi_{E}(0) \rangle $.

 The total Hamiltonian reads $ H=H_0+H_E+V,$ with $H_0,$ $H_E$ and $V$ representing the system, environment and interaction terms, respectively.
 As a result of the interaction, 
 system and environment become entangled at time $t\!:$ 
 \begin{equation}
 \label{t}
 | \Psi (t) \rangle = \frac {1} {\sqrt{2}} \left( | \psi_1(t)  \rangle \otimes | \psi_{E}^{(1)}(t) \rangle + | \psi_2(t) \rangle  \otimes | \psi_{E}^{(2)}(t) \rangle\right),
 \end{equation}
 and the  influence of the environment on the system coherence is captured by the complex amplitude $\langle \psi_{E}^{(2)}(t)| \psi_{E}^{(1)}(t) \rangle.$ 
 As in the previous section, we assume that the leading-order perturbation is of second-order in the interaction $V(\mathbf{r})$, i.e. the initial state $| \Psi (0) \rangle$ is such that $\langle  \Psi (0) |   V(\mathbf{r})  | \Psi (0) \rangle = 0$ at any position $\mathbf{r}.$ 
 The resulting influence of the environment  can then be 
 compared with the expression for 
the complex influence functional $S_{\rm IF}[ \mathbf{r}_1(t),\mathbf{r}_2(t)]$ derived from CTP integrals in the previous section:
\begin{eqnarray}
\label{eq:environment quantum states product}
e^{i S_{\rm IF}[ \mathbf{r}_1(t),\mathbf{r}_2(t)]/\hbar} &= &\langle \psi_{E}^{(2)}(t)| \psi_{E}^{(1)}(t) \rangle\nonumber \\
   &= &
  \langle \psi_{E}(0)  |  \widetilde{\mathcal{T}} e^{  \frac {i} {\hbar} \int_0^{\Delta t} dt  {\widetilde{V}}(r_2(t))}   \label{ptfinal}   \\
 && \times  \mathcal{T} e^{ - \frac {i} {\hbar} \int_0^{\Delta t} dt'  {\widetilde{V}}(r_1(t'))} \! |  \psi_{E}(0) \rangle, \nonumber
 \end{eqnarray}
where $ \mathcal{T}$ and $\widetilde{\mathcal{T}}$ denote respectively the time ordering and anti-time ordering operators.
One retrieves  the local influence functional (\ref{eq:phase local standard perturbation})   by
expanding each evolution operator in (\ref{ptfinal}) separately to second order, whereas
the non-local functional 
(\ref{eq:phase nonlocal standard perturbation})
is obtained by taking the product of first-order terms in 
(\ref{ptfinal}).

This approach  provides a complementary interpretation of the influence functionals $S_{\rm L}$ and 
$S_{\rm NL}$ in terms of the environment state vectors appearing in Eqs.~(\ref{0}) and (\ref{t}). 
We first develop the local decoherence functional by taking the imaginary part of (\ref{eq:phase local standard perturbation}) and using the identity 
${\rm Re}[ \langle A B \rangle ] = \frac 1 2 \langle \{ A, B   \} \rangle   $ for generic operators $A,B.$ We then  perform 
the variable change $t \leftrightarrow t'$ in the integrals involving anti-time ordered terms to find 
\begin{equation}
S_{\rm L}^{\rm Dec}[\mathbf{r}_1,\mathbf{r}_2] \! \ \! = \!  \!    \frac {1} {2 \hbar} \sum_{k=1,2} \int _{0}^{\Delta t} dt \int _{0}^{\Delta t} dt'    \langle {\widetilde{V}}(r_k(t))   {\widetilde{V}}
(r_k(t')) \rangle.
\end{equation}
We insert this expression into Eq.~(\ref{eq:definition stationary decoherence}) and use the completeness relation over 
the Hilbert space associated with the environment to find the local decoherence rate~\footnote{ The prefactor $\frac 1 2$ in 
(\ref{eq:local decoherence rate}) results from taking identical amplitudes for 
 the two system states $| \psi_1(0) \rangle $ and $| \psi_2(0) \rangle$ in the initial product state (\ref{0}).}
\begin{equation}
\Gamma_{\rm L} \! = \!  \lim_{\Delta t \rightarrow + \infty} \frac {1} {2 \Delta t}  \sum_{k=1,2}\sum_{| i_E \rangle } \left|  \langle i_E |  \frac {-i} \hbar \int_0^{\Delta t} \! \! dt {\tilde V}(r_k(t)) | \psi_E(0)  \rangle \right |^2 \label{eq:local decoherence rate}
\end{equation}
The sum can be naturally restricted to the environment quantum states $| i_E \rangle $ coupled to the initial state $ | \psi_{E}(0)  \rangle$ 
through the interaction potential $\widetilde{V}(\mathbf{r},t)$. Physically, the local decoherence rate appears as the average probability per unit time, to leading order in the interaction potential, that the initial state of the environment suffers a transition. Thus, the local decoherence rate is directly
 connected to the total transition rate from the initial environment state.

The nonlocal decoherence rate is instead related to the quality of the information flowing from the system into the environment. From 
Eq.~(\ref{eq:phase nonlocal standard perturbation}), the nonlocal influence functional $S_{\rm NL}$ can be expressed 
in terms of the overlap between the environment state perturbations when following each of the two system's paths:
\begin{equation}
\label{NLoverlap}
S_{\rm NL}[\mathbf{r}_1,\mathbf{r}_2]   =  - i\hbar \langle \psi_E^{(1),1}(\Delta t) | \psi_E^{(2),1}(\Delta t) \rangle,
\end{equation}
 with 
\begin{equation}
 | \psi_E^{(k),1}(\Delta t) \rangle =  -\frac {i} {\hbar} \int_{0}^{\Delta t} dt  \widetilde{V}(r_k(t)) | \psi_{E}(0) \rangle \label{eq:perturbation to the environment state}
 \end{equation}
representing 
the first-order perturbation  in the interaction picture when taking path $[{\bf r}_k(t)].$ 
In particular,
the nonlocal decoherence rate vanishes when the propagation of the system along the  paths $[\mathbf{r}_1(t)]$ and $[\mathbf{r}_2(t)]$ leads  to
orthogonal perturbations of the initial environment state. In this case, the interaction produces a perfect record of the system quantum state in the environment. 

More generally,
the nonlocal decoherence rate~(\ref{eq:definition stationary decoherence}) is proportional to the real part of the complex overlap amplitude,
\begin{equation}
\Gamma_{\rm NL}= - \lim_{\Delta t \rightarrow + \infty} \frac {1} {\Delta t} {\rm Re} \left[ \langle \psi_E^{(1),1}(\Delta t) | \psi_E^{(2),1}(\Delta t) \rangle \right] \,, \label{eq:expression non local decoherence rate}
\end{equation}
and thus may be positive or negative in principle. 
Nevertheless,
$\Gamma_{\rm NL}$ is certainly negative when the paths $[\mathbf{r}_1(t)]$ and $[\mathbf{r}_2(t)]$ become arbitrarily close to one another, since 
 the amplitude $ \langle \psi_E^{(1),1}(\Delta t) | \psi_E^{(2),1}(\Delta t) \rangle $ tends by continuity to $|| \, | \psi_E^{(1),1}(\Delta t) \rangle ||^2 = || \, | \psi_E^{(2),1}(\Delta t) \rangle ||^2 $ in this case.  

We can also discuss this limit by looking at the total decoherence rate  $\Gamma= \Gamma_{\rm L} +\Gamma_{\rm NL}.$
From Eqs.~(\ref{eq:local decoherence rate},\ref{eq:perturbation to the environment state},\ref{eq:expression non local decoherence rate}), one  shows that 
\begin{equation}\label{got}
\Gamma= \lim_{\Delta t \rightarrow +\infty} \frac {1} {\Delta t} \langle \psi_E^{(-)}(\Delta t)   |  \psi_E^{(-)}(\Delta t)  \rangle,
\end{equation}
where 
$  |  \psi_E^{(-)}(\Delta t)  \rangle = \frac {1} {\sqrt{2}} \left( | \psi_E^{(2),1}(\Delta t) \rangle  - | \psi_E^{(1),1}(\Delta t) \rangle \right).$
Eq.~(\ref{got}) resembles the result of Ref.~\cite{Breuer01}
expressing  the decoherence by emission of electromagnetic radiation 
 in terms of a current difference.

 When the paths $[\mathbf{r}_1(t)]$ and $[\mathbf{r}_2(t)]$ become arbitrarily close to one another, so do the perturbed states $ | \psi_E^{(1),1}(\Delta t) \rangle$ and $ | \psi_E^{(2),1}(\Delta t) \rangle $, yielding a vanishing total decoherence rate from (\ref{got}). 
 This results from a fine-tuning between a positive local decoherence rate $\Gamma_{\rm L}$ and a negative nonlocal decoherence rate $\Gamma_{\rm NL}.$ The latter may be interpreted as a recoherence of the system, reflecting that the environment 
 acquires imperfect which-way information about the system. While the local decoherence rate $\Gamma_{\rm L}$ guarantees the positivity of the total decoherence rate $\Gamma,$  only the nonlocal contribution $\Gamma_{\rm NL}$ captures the distinguishability between the two possible ``footprints'' left by the system on the environment when going along path $[\mathbf{r}_1(t)]$ or path $[\mathbf{r}_2(t)].$ 
 This term
 only survives if the environment has a sufficiently long memory time and in that sense
 is reminiscent of non-Markovian processes in open quantum systems~\cite{BreuerRMP16}.

\section{Decoherence by spontaneous emission and superradiance-like interference}

 We now illustrate the general arguments discussed above on a concrete example, which is an idealization of atom interferometry within an optical trap. Specifically, we consider  the spontaneous emission of a single excited atom prepared in a coherent superposition of two wave packets located in different wells of an external potential, as illustrated in the inset of Fig.~2. 
In the spirit of Ref.~\cite{Stern90}, our formalism allows us to interpret the decoherence either as resulting from a random phase, associated to the influence functional developed in Secs.~II and III, or in terms of entanglement with the dipole and electromagnetic field, discussed in Sec. IV. 
We show that the decoherence rate oscillates as a function of the distance $a$ between the wavepackets.  These oscillations, resulting from quantum interferences in the transition amplitudes,
enable us to build an analogy with superradiance. This is remarkable since our system contains only a single atom, 
while superradiance~\cite{Dicke54} is in essence a collective phenomenon. 
 
A closely related experiment~\cite{Chapman95} -- an atom interferometer undergoing photon scattering -- reported oscillations of the interference fringe contrast as a function of the distance between the arms. Other studies focused on the decoherence of the atomic internal degree of freedom~\cite{Eberly04,Ficek08} or on
 the influence of a largely spread center-of-mass wavefunction~\cite{Paul95}. 
 
 We consider an atom prepared in a coherent superposition state  of the form  $| \psi \rangle = \frac {1} {\sqrt{2}} \left( | \psi_+ \rangle + e^{i \varphi} | \psi_- \rangle \right),$  where $| \psi_{\pm} \rangle$ correspond to the trapped wavepackets in the two wells (tunneling neglected) located at $\mathbf{r}= \pm \frac 1 2 \mathbf{a}$ (see inset of Fig.~2) assumed distant enough so that $\langle \psi_- | \psi_+ \rangle=0$. We also assume that such superposition may be produced by an atomic beam-splitter leaving unaffected the atomic internal state  -- taken as an excited eigenstate $| e \rangle$ --, and that the two wave-functions are recombined coherently at a later time.

We apply Eqs.~(\ref{eq:phase local standard perturbation},\ref{eq:phase nonlocal standard perturbation}) 
to the case of a dipolar interaction Hamiltonian (in the interaction picture)
 $ {\tilde V}(\mathbf{r},t)= -  {\mathbf{d}}(t) \cdot  {\mathbf{E}}(\mathbf{r},t).$
 The 
 decoherence rates then involve  bilinear correlations in both dipole and electric field operators. Our first step is then to analyze such correlations. The dipole correlation reads  $\langle e | d_i(t) d_j(t')| e \rangle=\frac 1 3  \delta_{ij} \sum_{s} |\mathbf{d}_{s}|^2 e^{i \omega_{e s} \tau} $, the sum being performed over all possible internal
 atomic states $s$ with the Bohr frequencies $ \omega_{e s}= \frac 1 \hbar ( E_e -E_s) $
and $\mathbf{d}_{s}=\langle e | \mathbf{d}(0)| s \rangle.$
  The vacuum electric field correlation is well known~\cite{HeitlerBook}: $\langle 0 | E_i(\mathbf{r},t)E_j(\mathbf{r}',t') | 0 \rangle=\int d^3 \mathbf{k} \sum\limits_{\lambda}\left(\frac{ \hbar\omega}{(2 \pi)^3}\right)e^{i\mathbf{k} \cdot (\mathbf{r}-\mathbf{r}')- i \omega (t-t')}(\boldsymbol{\varepsilon}_{\mathbf{k}\lambda})_i(\boldsymbol{\varepsilon}_{\mathbf{k}\lambda})_j \,$ 
  with $\omega = k/c$ and 
  $\boldsymbol{\varepsilon}_{\mathbf{k}\lambda}$ representing a transverse unit vector associated to polarization $\lambda.$
The  decoherence rates  read ($\alpha=\rm{L,NL}$):
   \begin{equation}
   \label{eq:integral representation decoherence rates}
   \Gamma_{\alpha} \! = \! \! -     \frac 1 3 \sum_{s} |\mathbf{d}_{s}|^2 \! \int \!    d^3\mathbf{k} \left( \! \frac {\omega} {2  \pi^2 \hbar} \! \right) \! \!\int_{-\infty}^{+\infty} \! \! d \tau    e^{   i (\omega_{es}-\omega) \tau} \xi_{\alpha}
   \end{equation}
    where
   $
   \xi_{\rm{L}}= 1 $ and 
  $ \xi_{\rm{NL}}=- e^{i \mathbf{k} \cdot \mathbf{a} }.$

        As one may expect intuitively, the decoherence rates are closely related to the
      spontaneous emission rate   $ \gamma = \frac {1}{\Delta t}  \sum_s \int d^3 \mathbf{k} \left| \langle 1_{\mathbf{k}} | \otimes \langle s | \left(- \frac i \hbar \int_0^{\Delta t} dt  {V}(\mathbf{r},t) \right) | e \rangle \otimes | 0 \rangle \right|^2. $ 
 From Eq.~(\ref{eq:integral representation decoherence rates}), one derives that the local decoherence rate is indeed the half-sum of the spontaneous emission rates in each well. Since those are  identical and independent of the center-of-mass
  position, one finds $\Gamma_{\rm L}= \gamma$ for the local decoherence rate. 
     The nonlocal 
    decoherence rate 
 is obtained from Eq.~(\ref{eq:integral representation decoherence rates})  as   
 \begin{equation}
 \Gamma_{\rm NL} = -\sum_s \Gamma_{ es}\,{\rm sinc} \left(2\pi \frac {a} {\lambda_{es}} \right) ,
 \end{equation}
  where the  ${\rm sinc}(x) = \sin(x) / x$ and $\Gamma_{ es } = \omega_{es}^3 | \mathbf{d}_s |^2/ (6 \pi \hbar c^2)$ represents the
      spontaneous  transition rate  from level $e$ to level $s,$ associated to the transition wavelength $\lambda_{es}=2\pi  c/\omega_{es}.$ 
In Fig.~\ref{fig:decoherence rates}, we plot the total decoherence rate $\Gamma =  \Gamma_{\rm L}+\Gamma_{\rm NL}$
 as a function of the distance $a,$ for the case of a two-level atom (in this case $\gamma=\Gamma_{es}.$.  

 The  local and nonlocal decoherence rates only differ through the presence of the interference factor $ \xi_{\rm{NL}}=-e^{i \mathbf{k} \cdot \mathbf{a}}$  within the integral~(\ref{eq:integral representation decoherence rates}) for the latter, 
 which originates from  the two separate positions $\mathbf{r}_{\pm}=\pm \frac 1 2 \mathbf{a}$ taken in the r.-h.-s. of 
 Eq.~ (\ref{eq:phase nonlocal standard perturbation})
 (see  also Fig.~1b). 
 If the well separation is such that $a \ll \min_s \{ \lambda_{es} \},$ the interference factor 
  is then $ \xi_{\rm{NL}}\approx  -\xi_{\rm{L}}=-1,$ and the nonlocal recoherence effect  cancels the local decoherence, as illustrated by Fig.~2. 
  Physically, the wavelength of the photons emitted by the excited atom is too large for the two wells to be resolved within the diffraction limit. 
  Therefore, an observer detecting a spontaneously emitted photon could not tell from which well the radiation was emitted, and this lack of information is at the origin of the nonlocal recoherence. 
  
  In the opposite limit  $a \gg \max_s \{ \lambda_{es} \},$ 
   the nonlocal rate $\Gamma_{\rm NL}$ becomes negligible
   since it is given by the integration of a fast oscillating complex exponential. 
    Thus, the total decoherence rate saturates in the long-distance limit to the value fixed by the spontaneous emission rate, 
  which is represented by a horizontal dashed line in Fig.~2.  In this regime, increasing the distance $a$ between the wells will not  improve  the which-path information contained in each emitted photon, which already reveals almost certainly which well is occupied by the atom  (see also Ref.~\cite{Anglin97} for a related discussion). Alternatively, this can also be interpreted as follows: in the regime $a \gg \max_s \{ \lambda_{es} \},$ we have $a/c \gg 1/\omega_{\rm max}$ ($\omega_{\rm max}$ representing the largest Bohr frequency), so the 
    time-of-flight between wells  is much longer than the characteristic time of the atomic dipole correlation function, and then the coupling between forward and backward histories illustrated by Fig.~1b vanishes.   
   In other words,  the nonlocal decoherence rate contributes significantly only if the two atomic wavepackets  are closer than a characteristic length scale 
given by  the memory time of the environment times the propagation velocity of the interaction. 
  In appendix B, we study another example where the quality of which-path information 
saturates, for the case where this information is  
encoded in the frequency of the emitted photon.

The most interesting regime corresponds to intermediate values for the distance
 between the wavepackets, 
 for which
 the decoherence rate displays oscillations as shown in Fig.~\ref{fig:decoherence rates}. 
 Those oscillations are in contradiction with  the naive expectation that the decoherence rate should be a monotonic increasing function of the separation. 
 They may be interpreted in terms of  interference effects from photons co-produced by each atomic wavepacket in the superposition.  
  For certain distances, decoherence can be faster than in the case of infinite separation, where the which-way information would in principle be the most accurate. This is in close analogy with the behavior of the photon emission rate in the case of superradiance \cite{Dicke54} with two atoms separated by a distance of the order of the transition wavelength. 
  Note that in the single-atom case considered here we have interference of quantum amplitudes associated to information on the atomic position, but not in the photon emission rate, which 
  would be present only in the case of wavepacket overlap.

\begin{figure}[htbp]
\begin{center}
\includegraphics[width=8 cm]{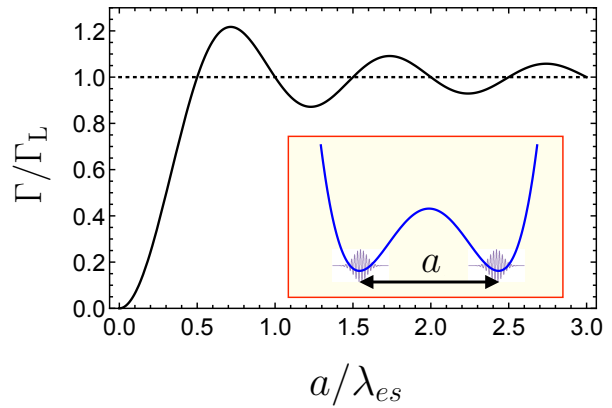}
\end{center}  \caption{(color
  online). Decoherence rate  variation with the distance $a$ between the two wells. 
  The total rate $\Gamma$
   is expressed in units of the local contribution $\Gamma_{\rm L},$ which does not depend on $a,$
 whereas the distance is expressed in units of the transition wavelength $\lambda_{es}.$ For simplicity, we have considered a two-level atom
  in the excited state. 
  The quantum state of the atom center-of-mass is initially a coherent superposition of localized wave packets in a
  double-well potential, as illustrated by the inset. }
\label{fig:decoherence rates}
\end{figure} 

\section{Conclusion}

In conclusion, we have  shown that quantum interferences may enhance or reduce the decoherence of the center-of-mass position of a single excited atom. 
As a consequence, this decoherence can be fastest at a finite separation.
The resulting oscillations suggest an analogy with superradiance, with the photon emission rate replaced by the decoherence rate. 
Instead of multiple atoms, here we 
have multiple wave packets giving  rise to different quantum paths. 
 This enhancement of decoherence does not result from an increased spontaneous emission rate, but rather from interferences involving the quantum states of the environment associated to distinct paths. 

The quantum interferences observed in the decoherence rate are entirely related to the contribution of the nonlocal functional, which results from the coupling between forward and backward histories.
This contribution, often neglected in the dynamics of open quantum systems, captures the quality of the 
which-way information flowing from the system into the environment. It is actually essential to estimate the decoherence of a system interacting with a non-Markovian environment.

\acknowledgements

 The authors are grateful to Robin Kaiser, Ryan Behunin  and Claudio Ccapa-Ttira for enlightening suggestions and discussions.
This work was supported by the Brazilian agencies CNPq and  FAPERJ.

\appendix

\section{Explicit derivation of the influence functional}

In this appendix we present the CTP derivation of Eqs.~(\ref{eq:phase local standard perturbation}) and (\ref{eq:phase nonlocal standard perturbation}) for the local and nonlocal influence functionals, respectively. 
For the sake of
clarity, from now on we identify the operators with a hat.
To simplify the notation, we start with a simple system composed only of two d.o.f.s $X$ and $q$. For convenience, we recall the formal expression of the influence functional, obtained by integrating over the d.o.f.s $X$~\cite{Feynman1963,CalzettaHuBook,MiltonBook15}:
\begin{eqnarray}
e^{\frac{i}{\hbar}S_{\rm IF}[q,q']} &= &\int dX_f\int_{\rm CTP}^{X_f,X_f} \! \! \mathcal{D}X e^{\frac{i}{\hbar}(S_X[X]-S_X[X'])} \nonumber \\
& \: & \qquad \times e^{\frac i \hbar (S_{\rm int}[X,q]-S_{\rm int}[X',q'])} \, , \label{sif}
\end{eqnarray}
where $S_X[X]$ stands for the free action involving the d.o.f. $X$ only and
$S_{\rm int}[X,q]$ is the interaction term, which couples $X$ and $q.$ 
We  require the interaction to be local:  
 $S_{\rm int}[X,q]=\int_{0}^{T}dt \,V(X(t),q(t)),$
 where $V$ is a generic potential.
 This action is identical to $(\ref{eq:definition Sint})$ except for the position-dependence of the potential, which we omit
  for the time being to render the presentation simpler.
  
 We also remind the definition of the CTP integration $(\ref{eq:CTP notation})$ 
 \begin{equation}
 \int_{\rm CTP}^{X_f,X'_f}  \! \! \!   \! \! \!   \! \! \!  \mathcal D X \!  = \!  \int \! d X_0 d X'_0 \! \! \int_{X(0)=X_0}^{X(\Delta t)=X_f}  \! \! \!  \! \! \! \!  \! \! \!  \mathcal D X \! \int_{X'(0)=X'_0}^{X'(\Delta t)=X'_f}  \! \! \! \!  \! \! \!   \! \! \!   \! \! \!   \mathcal D X'   \langle X_0 | 
 \rho (0) | X_0' \rangle
\end{equation}
 
We assume that the coupling can be treated perturbatively, and that the leading order term is of second order. 
The  influence functionals are then written as
\begin{eqnarray}
\label{eq:influence actions appendix}
S^{\rm IF}_{\:^{\rm L}}[q,q']&=&\frac{1}{4\hbar} \int dX_f\int_{CTP}^{X_f,X_f}\mathcal{D}X e^{\frac{i}{\hbar}(S_X[X]-S_X[X'])}  \nonumber \\
 & \: & \qquad \qquad  \times  (S_{\rm int}^2[X,q]+S_{\rm int}^2[X',q']) 
\end{eqnarray}
\begin{eqnarray}
S^{\rm IF}_{\:^{\rm NL}}[q,q'] & = & - \frac{1}{2\hbar} \int dX_f\int_{CTP}^{X_f,X_f}\mathcal{D}X e^{\frac{i}{\hbar}(S_X[X]-S_X[X'])} \nonumber \\
 & \: & \qquad \qquad  \times \left( S_{\rm int}[X,q]S_{\rm int}[X',q'] \right)
\end{eqnarray}

Let us develop in detail the expression of the nonlocal influence functional (the local one can be obtained following similar steps). A key point is that 
the integrals~(\ref{eq:influence actions appendix})  can be performed separately on the paths $X$ and $X'$.
For example, we perform the integral over the path $X\!:$ 
\begin{equation}\label{Iq}
\mathcal{I}[q]=\int_{X_0}^{X_f}\mathcal{D}X e^{\frac{i}{\hbar}S_X[X]}S_{\rm int}[X,q] 
\end{equation}
We may write (\ref{Iq})  as 
\begin{eqnarray}
\mathcal{I}[q]&=&\int_0^{\Delta t} dt\int dX(t) \left(\int_{X_0}^{X(t)}\mathcal{D}X e^{\frac{i}{\hbar}S_X[X]}\right) 
V(X(t),q(t)) \nonumber \\
& \:&  \qquad \qquad \qquad \qquad \qquad  \times \left(\int_{X(t)}^{X_f}\mathcal{D}X e^{\frac{i}{\hbar}S_X[X]}\right)
\end{eqnarray}

We now proceed to evaluate this quantity with the help of the propagator
\begin{eqnarray}
\langle X_2 |   \hat{U}(t,t') | X_1 \rangle & = &  \int_{X(t')=X_1}^{X(t)=X_2} \mathcal{D}X \: e^{\frac{i}{\hbar}S_X[X]} 
\end{eqnarray} 
where $\hat{U}(t,t')$ is the evolution operator. 
As in Ref.~\cite{CalzettaHuBook}, we  note that
the potential $V(X(t),q(t))$ is an eigenvalue associated with the vector $| X(t) \rangle.$ We write\\
\begin{eqnarray}
\langle X_f |\hat{U}(\Delta t,t)| X(t) \rangle
V(X(t),q(t))\cr\cr
=\langle X_f |\hat{U}(\Delta t,t)V(\hat{X},q(t))| X(t)\rangle
\end{eqnarray}
so that one obtains
\begin{eqnarray}
\label{Iqfinal}
\mathcal{I}[q]
=\int_0^{\Delta t} dt\langle X_f | \hat{U}(\Delta t,t)V(\hat{X},q(t))\hat{U}(t,0)| X_0 \rangle.
\end{eqnarray}
The path integral over $X'$, 
\begin{equation}
\label{Iprime}
\mathcal{I}'[q']=\int_{X_0'}^{X_f'} \mathcal{D}X' e^{-\frac{i}{\hbar}S_X[X']}  S_{\rm int}[X',q']
\end{equation}
 can be expressed similarly as 
\begin{eqnarray}
\mathcal{I}'[q']
=\int_0^{\Delta t} dt'\langle X'_0 | \hat{U}(0,t')V(\hat{X},q(t'))\hat{U}(t',\Delta t)| X_f \rangle \, \nonumber \\ \label{Iprimefinal}
\end{eqnarray}
Note the backward evolution, which results  from
taking the action with the $-1$ prefactor in (\ref{Iprime}).
 Using the expressions
 (\ref{Iqfinal}) and 
 (\ref{Iprimefinal})
  for $\mathcal{I}[q]$ and $\mathcal{I}'[q']$ as well as the completeness relation, one obtains
\begin{eqnarray}
\label{eq: nonlocal appendix final} 
S_{\rm NL}&=&-\frac{i}{\hbar}\int_{0}^{\Delta t}dt\int_{0}^{\Delta t}dt' \int dX_f    \\
& &  \langle X_f|V(\hat{X}(t'),q(t'))  V(\hat{X}(t),q(t))         \hat{\rho}(0)|X_f\rangle\, , \nonumber \\
& = & -\frac{i}{\hbar}\int_{0}^{\Delta t}dt\int_{0}^{\Delta t}dt'  \langle V(\hat{X}(t'),t') V(\hat{X}(t),t) \rangle \nonumber
\end{eqnarray}
We have introduced the interaction picture operators $ \hat{X}(t)= U(0,t) \hat{X} U(t,0)$ and used the cyclic property of the trace as well as the 
group property of the evolution operator $\hat{U}(t_3,t_2)\hat{U}(t_2,t_1)=\hat{U}(t_3,t_1)$. 

Finally, we now include the possibility of $X$ to be a field so that the interaction
of $X$ with $q$ becomes dependent of the center-of-mass position
of the system  and given by (\ref{eq:definition Sint})
in Sec.~III.
We assume the wavepacket width to be negligible (for a detailed account of the width see Ref.~\cite{Impens14}). In this case, the derivation
provided above remains essentially the same, 
except for the replacement of  ${\hat X}(t)$
and ${\hat X}'(t')$
 by ${\hat X}(\mathbf{r}_1(t),t)$ and
${\hat X}'(\mathbf{r}_2(t'),t'),$ respectively. {With such replacements, one obtains a functional depending on two paths, i.e. Eq.~(\ref{eq: nonlocal appendix final}) becomes}
\begin{eqnarray}
\label{eq: nonlocal}
S_{\rm NL}[\mathbf{r}_1(t),\mathbf{r}_2(t)] \! = \!  -\frac{i}{\hbar} \! \int_{0}^{\Delta t}\! \! \! \! \int_{0}^{\Delta t}\! \! dtdt' \! \langle V(\hat{X}(\mathbf{r}_2(t'),t') V(\hat{X}(\mathbf{r}_1(t),t)\rangle  \nonumber
\end{eqnarray}
which is 
equivalent to
 Eq.~(\ref{eq:phase local standard perturbation}) . The local influence functional can be derived by following the same steps.

\section{Quality of information in the time domain}


In this appendix, we discuss another example showing that the nonlocal decoherence contribution represents the quality of which-way information.    We consider the evolution of the system during a finite time $\Delta t$ much greater than the decay time of the dipole and field correlations. One expects that only nearly resonant processes may contribute significantly to the decoherence rate over the considered time scale $\Delta t$. This corresponds to the emission of photons with frequencies in the vicinity of the  atomic Bohr frequencies $\omega_{e s}.$ Below, we show that a small mismatch of the Bohr frequencies $\omega_{e s}$ in the two wells affects the local and nonlocal decoherence rates in very different ways. \\

From now on, we assume that the atomic levels are light-shifted differently in each well, yielding the distinct Bohr frequencies $\omega_{es}^{\pm}.$ To simplify the discussion, we assume that the
values of the atomic transition matrix elements   $|\mathbf{d}_s|^2$  are not significantly modified. To obtain the local and nonlocal decoherence rates, one uses again  Eqs.~(\ref{eq:local decoherence rate},\ref{eq:perturbation to the environment state},\ref{eq:expression non local decoherence rate}).

     The local decoherence rate involves two successive transitions through intermediate states of identical energies. One obtains $\Gamma_{\rm L}=\frac 1 2 (\Gamma_L^++\Gamma_L^-),$ where the rates $\Gamma_L^{+}$ and $\Gamma_L^-$ correspond to the local rate $\Gamma_{\rm L}$ as given by Eq.~(\ref{eq:integral representation decoherence rates}) up to a change of the Bohr frequencies from $\omega_{es}$ to $\omega_{es}^+$ and $\omega_{es}^-$ respectively. Typically, the change of the local decoherence rate is extremely small. In contrast, the nonlocal decoherence rate involves transitions through atomic states dressed differently by the light field. Precisely, the nonlocal decoherence rate involves the integration of oscillating functions 
    \begin{eqnarray}
   \Gamma_{\rm NL}  \! & = &  \!   -    \!  \! \! \!  \!    \frac {1} {3\Delta t}  
    \sum_s  |\mathbf{d}_{s}|^2 \int d^3 \mathbf{k} \left( \! \frac {\omega} {2  \pi^2 \hbar} \! \right)  e^{i \mathbf{k} \cdot \mathbf{a} }  \\
 & \times &    \int_0^ {\Delta t} \!\! d t_m e^{-  i  (\omega_{es}^{+}-\omega_{es}^{-} ) t_m } \int_{2t_m-2\Delta t}^{2 t_m} \!\!  d \tau   e^{-  \frac i 2 (\omega_{es}^{+}+ \omega_{es}^{-}) \tau}     \nonumber .
     \end{eqnarray}
     where we introduced the time variables $t_m=\frac 1 2 (t+t')$ and $\tau=t-t'$. The integral over $\tau$ can be extended to infinity and yields, together with the integral over $\mathbf{k}$, a finite result. The integral over the time $t_m$ fails to provide a term proportional to $\Delta t$ because of the oscillatory complex exponential arising from the light shift mismatch.     
  One then finds that 
  the nonlocal decoherence rate is significant only for short monitoring times or small frequency mismatches,
  $ |\omega_{es}^{+}- \omega_{es}^{-}| \stackrel{<}{\scriptstyle\sim} 1 / \Delta t.$ 
     
     Again, we may interpret this  result in terms of quality of information. We assume here that the two wells are close $a/\lambda_{es} \ll 1 $ for all levels $s$, so that 
     which-path
     information can only be stored in the frequency value  $ \omega_{es}^{+}$ or  $\omega_{es}^{-}.$
      By monitoring the system during $\Delta t$, 
      the frequency resolution is limited by $\delta \omega \simeq  1 /  \Delta t$.
     If $ |\omega_{es}^{+}- \omega_{es}^{-}| \stackrel{<}{\scriptstyle\sim} 1 / \Delta t,$ an observer detecting a photon emitted from one of the wells 
     would not be abe to resolve the frequency with the accuracy needed to 
   determine the location of the emitting atom. 
   Thus, the emission process carries which-way information of low quality in this case. 
   Consistently with our previous interpretation, the nonlocal decoherence term is significant and of negative sign, thus providing recoherence to the system. 
   On the other hand, if $|\omega_{es}^{+}- \omega_{es}^{-}|  \gg 1 / \Delta t,$  the nonlocal decoherence rate associated to the level $s$ vanishes. In this limit, the detection of a photon emitted from a transition $e \rightarrow s $ provides  perfect information on the atomic position given the frequency resolution  achievable during the monitoring time $\Delta t.$ 
   This is another example showing that the presence of a nonlocal stationary recoherence rate is tied to imperfections in the which-way information gained through the interaction process.\\

\bibliography{160714biblio_geral}{}

\end{document}